\documentclass[aps,pra,twocolumn,superscriptaddress,showpacs]{revtex4-2}
\usepackage{amsmath}
\usepackage[dvips,xdvi]{graphicx}
\usepackage{amssymb}
\usepackage{epsfig}
\newcommand{\ket}[1]{| #1 \rangle} 

\begin{document}
\title{Quantum to Classical Walk Transitions Tuned by Spontaneous Emissions}
\author{J. H. Clark}
\affiliation{Department of Physics, Oklahoma State University, Stillwater, Oklahoma 74078, USA}
\author{C. Groiseau}
\affiliation{Dodd-Walls Centre for Photonic and Quantum Technologies, New Zealand}
\affiliation{Department of Physics, University of Auckland, Auckland 1010, New Zealand}
\author{Z. N. Shaw}
\affiliation{Department of Physics, Oklahoma State University, Stillwater, Oklahoma 74078, USA}
\author{S. Dadras}
\affiliation{TOPTICA Photonics Inc., 5847 County Road 41, Farmington, New York 14424, USA}
\author{C. Binegar}
\affiliation{Department of Physics, Oklahoma State University, Stillwater, Oklahoma 74078, USA}
\author{S. Wimberger}
\email{Electronic address: sandromarcel.wimberger@unipr.it} \affiliation{Dipartimento di Scienze Matematiche, Fisiche e Informatiche, Università di Parma, Campus Universitario, Parco area delle Scienze 7/a, 43124 Parma, Italy}
\affiliation{INFN - Sezione di Milano-Bicocca, gruppo collegato di Parma, Parco Area delle Scienze 7/A, 43124 Parma, Italy}
\author{G. S. Summy}
\email{Electronic address: gil.summy1@gmail.com}
\author{Y. Liu}
\email{Electronic address: yingmei.liu@okstate.edu} \affiliation{Department of Physics, Oklahoma State University,
Stillwater, Oklahoma 74078, USA}
\date{\today}
\begin{abstract}
We have realized a quantum walk in momentum space with a rubidium spinor Bose-Einstein condensate by applying a periodic
kicking potential as a walk operator and a resonant microwave pulse as a coin toss operator. The generated quantum walks
appear to be stable for up to ten steps and then quickly transit to classical walks due to spontaneous emissions induced
by laser beams of the walk operator. We investigate these quantum to classical walk transitions by introducing
well-controlled spontaneous emissions with an external light source during quantum walks. Our findings demonstrate a
scheme to control the robustness of the quantum walks and can also be applied to other cold atom experiments involving
spontaneous emissions.
\end{abstract}

\maketitle
\section{Introduction}
Quantum walks (QWs) have been actively studied in many experimental systems, such as photons, lattice-confined atoms, and
trapped ions, since the first theoretical model was introduced in
1993~\cite{PhysicalimpQW,QWOplat2002,QWoptLat2005,QWtrappedion,QWscience2009,QW1st1993,QWBrillouin}. Possessing spin degrees of freedom,
spinor Bose-Einstein Condensates (BECs) have also been suggested as ideal candidates for QW implementation~\cite{QWpropBECfirst}. Two important
components of QWs are a walk operator to shift a walker in positions or momentum space and a coin toss operator to
determine the direction that the walker shifts in each step~\cite{QWintroreview}. In this work, a rubidium spinor BEC subjected to a series of periodic optical pulses,
which can be described as an atom-optics kicked rotor (AOKR), is utilized to create a QW in momentum
space~\cite{SiamakPRA,SiamakPRL,QauntumsearchSKW,SpontEmissPRA}. These periodic pulses construct one-dimensional (1D)
optical lattices and act as a walk operator in momentum space. Resonant microwave pulses, entangling two hyperfine spin
states, are the coin toss operator. In contrast to classical random walks with Gaussian distributions, QWs distribute
ballistically because atoms conducting QWs can be in a superposition state~\cite{PhysicalimpQW,SiamakPRA,SiamakPRL}. Other
advantages of QWs studied in this paper include hitting target points faster than classical walks, fast propagation, and
entanglement between internal and external degrees of freedom~\cite{PhysicalimpQW,QWPortugalbook}. QWs thus have many
proposed and realized applications in various research fields including quantum information, metrology, and topological
phenomena~\cite{QWtopology}.

In this work, we demonstrate that our quantum walk in momentum space can be stable for up to ten steps and then quickly
transit to classical walks due to spontaneous emission (SE) induced by the laser beams imprinting a momentum change. The
SE effects have been observed in our previous experiments and are pervasive in other experiments utilizing
AOKR~\cite{AOKRdeco2013,AOKRdecoherence}.

We investigate the SE-tuned quantum to classical walk transitions by introducing well-controlled SE with an
additional laser which does not interfere with the kick or shift laser of our QW. Those effects are manifold since SE acts as
projective measurement in the internal electronic spin degree of freedom of the atom. On the other hand, SE has the
twofold effect on the external center-of-mass degree of freedom of the atoms in our BEC: first it changes the
quasimomentum and hence the conditions of being in the QW or not, see~\cite{SiamakPRA,SiamakPRL}, and secondly, it biases
the QW towards the direction of the ground state into which the electronic degree is projected.
This is also contrary to previous experiments~\cite{AOKRdecoherence} with just one effective internal state in which SE only had an influence on the quantum-resonance condition and hence on the external degree of freedom. In our
experiments, the probability of a SE event and the induced decoherence appear to increase with the evolution of time,
i.e., with the number of steps in a QW. We also confirm the SE events lead to a biased momentum distribution, which agrees
well with our numerical simulations. Our findings demonstrate a scheme to control the robustness of quantum walks and
can also be applied to other cold atom experiments involving spontaneous emissions~\cite{Andersen1,Andersen2}.

\begin{figure*}[tb]
\includegraphics[width=150mm]{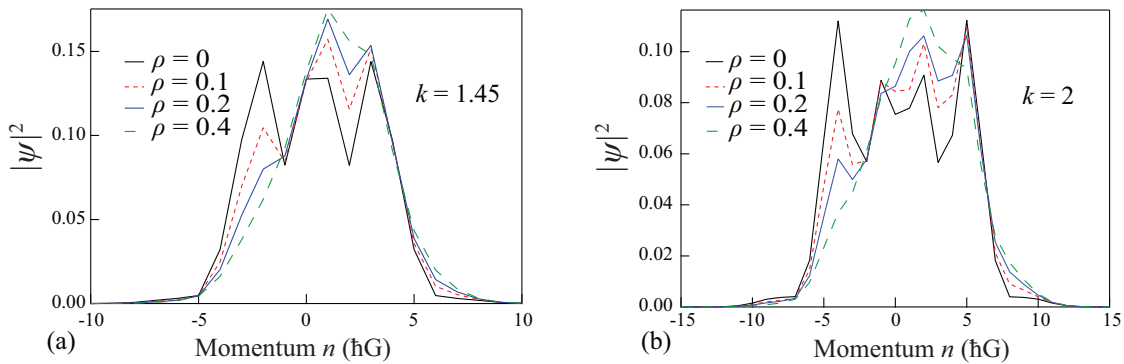}
\caption{Simulated momentum distributions of $5$-step quantum walks at two kicking strengths, $k=1.45$ (a) and $k=2.0$
(b), averaged over $1\times10^3$ trajectories with $\Delta_{\beta} = 0.025 \hbar G $ at various probabilities $\rho$ of SE events. Here $\Delta_{\beta}$ is the width of the
quasimomentum $\beta$. Note the increasing asymmetry as the probability $\rho$ increases.}
\label{superpositionplots}
\end{figure*}

\section{Theoretical Predictions}
Similar to our previous works, we describe each QW step with an operator $ \bf{\hat{U}}_{step}=\bf{\hat{T}}\bf{\hat{M}}
$~\cite{SiamakPRA,SiamakPRL}. A unitary walk operator $ \bf{\hat{T}} $ implemented by AOKR entangles the internal (i.e.,
spin) and external degrees of freedom, which leads to a momentum change of $ p_{m}= m\hbar
G$~\cite{SiamakPRA,SiamakPRL,QauntumsearchSKW,SpontEmissPRA}. Here $G$ is the wavevector of the 1D lattice, $\hbar$ is the
reduced Planck's constant, and $m$ is an integer number. The coin operator $ \bf{\hat{M}} $, created by a microwave pulse
resonant with the transition between $ \ket{F=1, m_{F}=0} $ and $ \ket{F=2, m_{F}=0} $ states of $^{87}$Rb atoms, produces
a superposition of these two internal states. We apply a controlled amount of SE during a QW sequence using an
independently controlled laser, which excites atoms from the $|F=2\rangle$ ground state resonantly to the $|F'=3\rangle$
excited state. The laser coupling $\Omega =\gamma\sqrt{\frac{I}{2I_s}}$ is small compared to the excited state hyperfine
splitting between $|F'=3\rangle$ and $|F'=2\rangle$, so that the $|F=2\rangle\rightarrow|F'=2\rangle$ transition can be
assumed to be too far detuned to create a significant population in $|F'=2\rangle$. Here $I$ is the intensity of the
laser, $I_s$ is the saturation intensity, and $\gamma$ is the decay rate~\cite{steck2001rubidium}. Due to selection rules,
the atom can only decay from $|F'=3\rangle$ back to $|F=2\rangle$, corresponding to a projection of the atom onto
$|F=2\rangle$. The SE pulse is long enough that we can assume the atom reaches the steady-state (the coin pulse should not
interfere with that) meaning that the effective SE-rate $\gamma_{\text{eff}}$ is given by the natural line width times the
steady-state population of $|F'=3\rangle$ as follows~\cite{steck2001rubidium},
\begin{equation}
    \gamma_{\text{eff}}=\frac{\gamma}{2}\frac{I/I_s}{1+I/I_s},
\end{equation}
from which we get the probability of an event per pulse
\begin{equation}
    \rho =\gamma_{\text{eff}}t_{\text{SE}}.
\end{equation}
Technically, $\gamma_{\text{eff}}$ and $\rho$ change during a single trajectory. The probability of the first decay overall
and in each further decay has to be scaled down by a factor 2 since the atom will be either exactly or close to an equal superposition
of the two ground states. So the given rate represents an upper limit, good only a couple of $\mu s$ after an event. We estimate $\rho\approx0.35$ for a SE power of $3~\mu \rm{W}$ for our experimental system, as elaborated in
Section-III.

Since the SE light is introduced 30~$\mu$s after the start of the coin, SE events will interrupt the coin pulse at
random times, the partial action of the coin operator in between two events of time delay $t$ is
\begin{equation}
    e^{i\frac{\pi t}{4 T}\hat \sigma_x}=\begin{bmatrix}
        \cos(\frac{\pi t}{4T}) & i\sin(\frac{\pi t}{4T}) \\
        i\sin(\frac{\pi t}{4T}) & \cos(\frac{\pi t}{4T})
    \end{bmatrix},
\end{equation}
where $T$ is the total length of the coin pulse.

This means that the state of the internal degree of freedom at the end of the coin sequence is only determined by the time
of the last SE event $t'\in[0.29,0.58]\times T$ and thus given by
\begin{equation}
|\psi\rangle= \cos[\frac{\pi (T-t')}{4T}]|2\rangle+ i\sin[\frac{\pi (T-t')}{4T}]|1\rangle.\label{SEcoin}
\end{equation}
Here $|1\rangle$ and $|2\rangle$ represent the two internal states, $ \ket{F=1, m_{F}=0} $ and $ \ket{F=2, m_{F}=0} $,
respectively. Eq.~\eqref{SEcoin} clearly shows that SE creates an imbalance in the internal state of the atoms towards
$|F=2\rangle$, which gets transferred to the populations and results in a biased momentum distribution (see our
simulations in Fig.~\ref{superpositionplots}).

Each SE event also affects the external degree of freedom by shifting the quasimomentum $\beta$ by a random amount.
Contrary to SE induced by the kicking beams~\cite{SpontEmissPRA}, the atom does not incur any recoil from the absorption
of a photon from the SE beam due to its perpendicular alignment to the walk axis.

In our simulations, we draw up to $3$ Poisson-distributed times and perform the partial coin operator from the largest time that is still inside the coin duration. We also add the corresponding amount of random recoil (here taken to be uniformly distributed). Typical simulation results for $5$-step quantum walks at two kicking
strengths $k$ are shown in Fig.~\ref{superpositionplots}, which clearly show transitions from quantum walks to classic
walks as the probability $\rho$ of SE events increases.\vspace{-1pc}
\begin{figure}[tb]
\includegraphics[width=85mm]{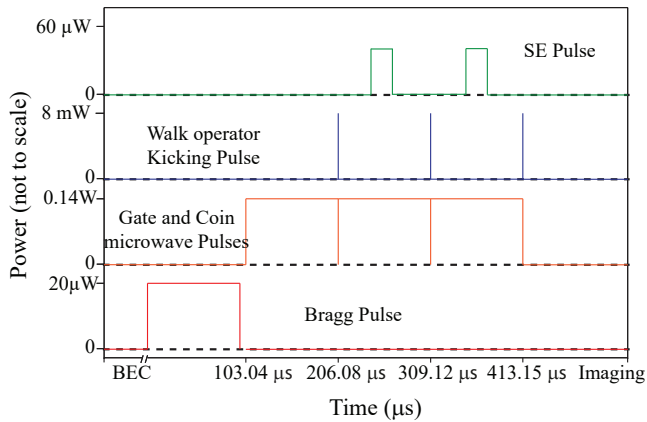}
\caption{Schematic diagram showing the sequence of various optical and microwave pulses used in our experiments. The time
duration of each SE pulse is $ 30~\mu$s. All axes are not to scale.}\label{schematic}
\end{figure}

\begin{figure*}[tb]
\includegraphics[width=170mm]{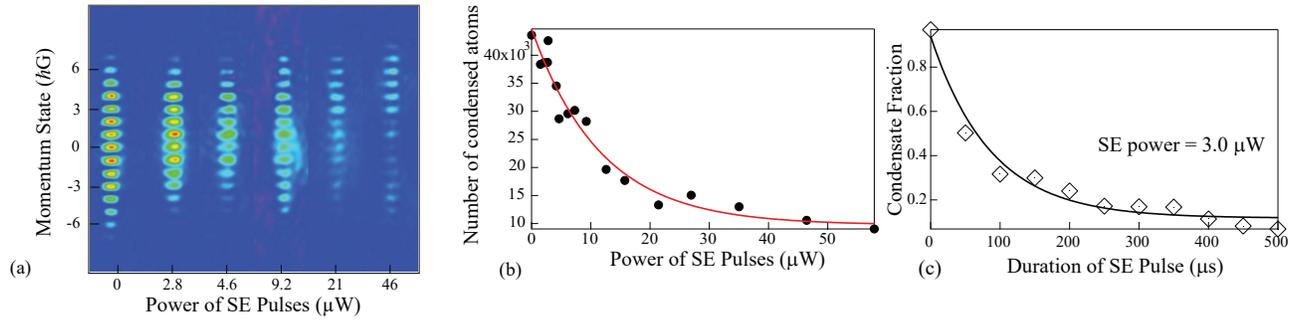}
\caption{(a) TOF images of a phase non-compensated five-step QW under various SE powers at the kicking strength $k=1.45$
and the SE pulse duration of 30~$\mu$s. (b) The number of condensed atoms versus the SE power in the noncompensated QWs shown in Panel~(a). (c) The condensate fraction versus SE pulse duration for a single BEC subjected to a pulse of SE light at SE = 3.0 $\mu W $. Solid lines in Panel~(b) and Panel~(c) are exponential fits.}\label{unbalanced}
\end{figure*}

\begin{figure*}[tb]
\includegraphics[width=153mm]{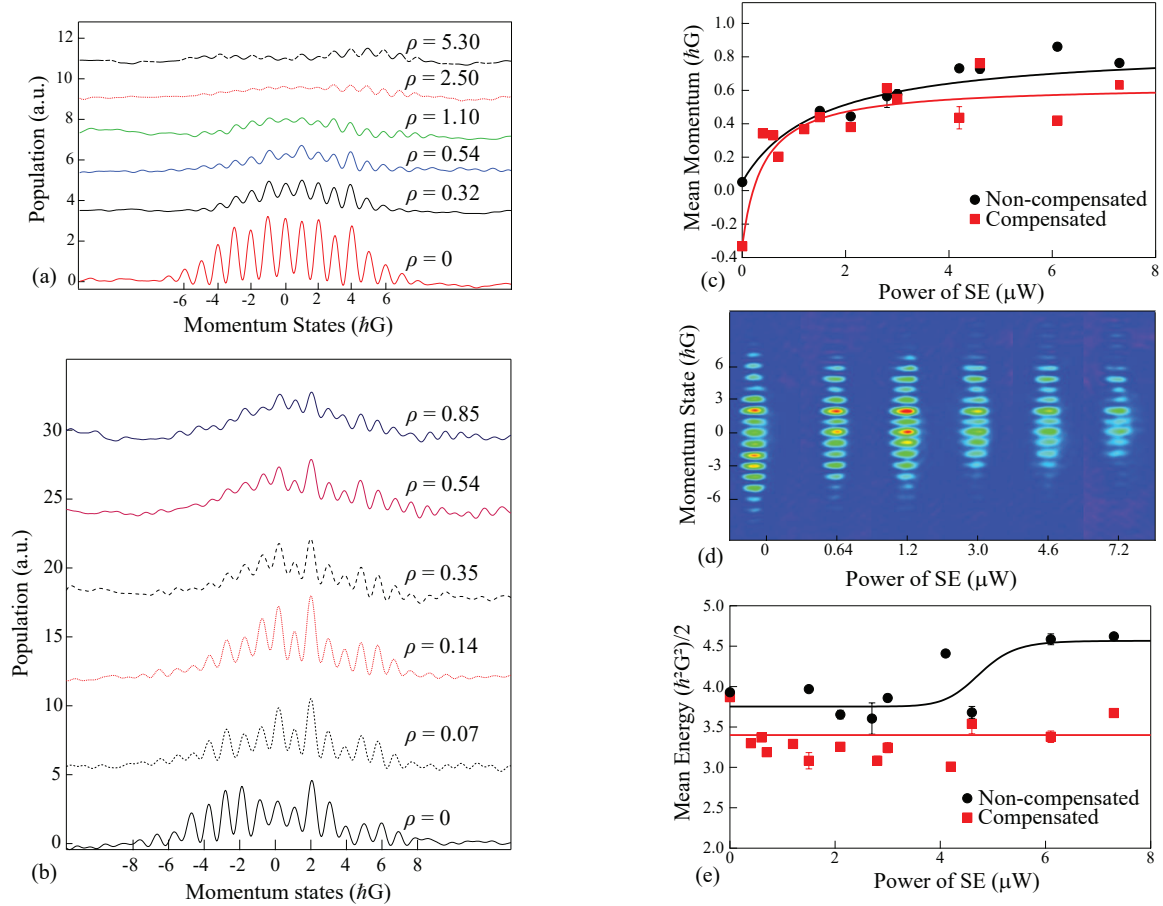}
\caption{Typical momentum distributions of (a) a phase non-compensated QW and (b) a phase compensated QW at various $\rho$ and at a fixed SE pulse duration of 30~$\mu$s (see text). Each
momentum distribution is shifted by a constant offset for visual clarity. (c) The mean momentum extracted from Panels (a)
and (b) as a function of the SE power. (d) Typical TOF images in the compensated QWs. (e) The mean energy extracted from
Panels (a) and (b) versus the SE power. Solid lines in Panel~(c) and Panel~(e) are fitting curves to guide the
eye.}\label{momentum}
\end{figure*}

\begin{figure*}[tb]
\includegraphics[width=153mm]{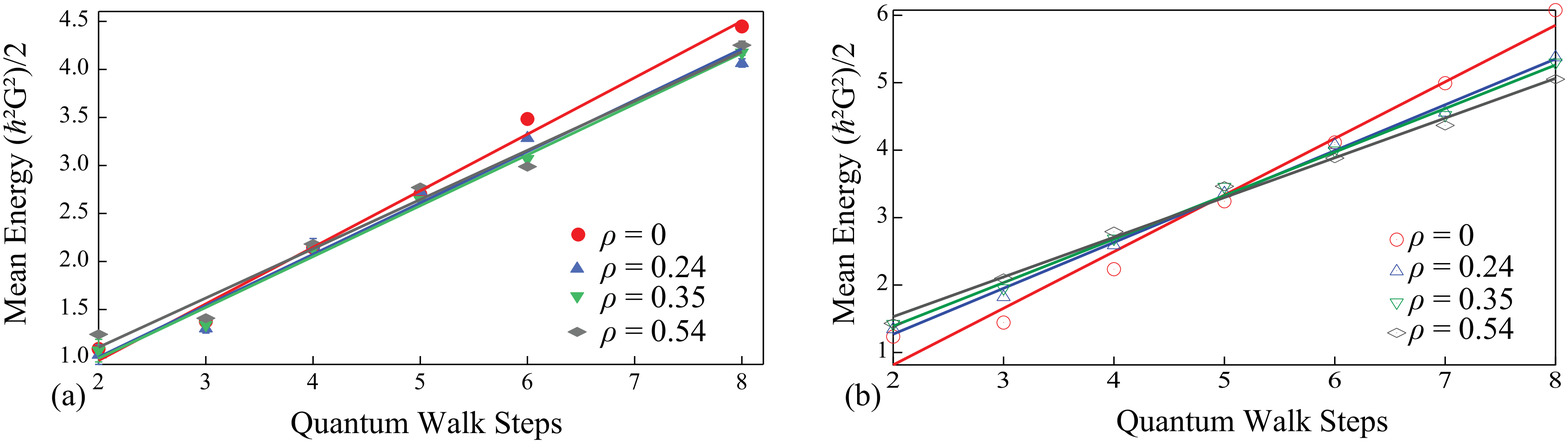}
\begin{center}
    (c)
    \begin{tabular}{ |c|| l| r| }
        \hline
        SE Probability & Experimental Rate & Theoretical Rate \\
        \hline\hline
        $ \rho = 0 $ & 0.58 & 0.84 \\
        $ \rho = 0.24 $ & 0.53 & 0.68 \\
        $ \rho = 0.35 $ & 0.52 & 0.64 \\
        $ \rho = 0.54 $ & 0.51 & 0.58 \\
        \hline
    \end{tabular}
\end{center}
\caption{(a) The extracted mean energy as a function of the QW steps at various probability $\rho$ of SE events and
$k=1.4$. The difference in the mean energy among various probability $\rho$ of SE events remains relatively constant at
lower number of steps but begins to decrease at six steps, indicating a transition from a QW to a classical walk. (b)
Theoretical predictions based on the experimental conditions for the data shown in Panel~(a). The solid lines in both
panels represent the linear fitting functions. (c) Rate $R$ extracted from Panel~(a) and Panel~(b) (see
text).}\label{meanenergy}
\end{figure*}

\section{Experimental Procedures}
Each experimental sequence starts with a BEC of approximately $4\times 10^4$ $^{87}$Rb atoms at the $\ket{F=1,m_{F}=0} $
state. The BEC is then subjected to Bragg, AOKR and microwave pulses. A schematic outlining of the pulse sequences is shown
in Fig.~\ref{schematic}. The AOKR and Bragg pulses are realized with the same two counter-propagating laser beams that
intercept on the BEC, although the Bragg pulse has a longer duration to drive the BEC into the state
$\ket{\psi_{0}}=1/\sqrt{2}(\ket{n=0}+e^{i\phi}\ket{n=1}) $~\cite{BraggRef,SiamakPRA}. We control populations of the two
internal states $|1\rangle$ and $|2\rangle$ using the microwave (coin toss) pulses resonant with the $|1\rangle$ to
$|2\rangle$ transition.

The standard QW of $i$ number of steps is created with a sequence of pulses described by the operator
$(\bf{\hat{U}}_{step})^{i}=[\bf{\hat{T}}\bf{\hat{M}}(\pi/2,-\pi/2)]^{i-1}[\bf{\hat{T}}\bf{\hat{M}}(\pi/2,\pi)] $. To
ensure that the QW is symmetric the first coin pulse in the sequence is a Hadamard gate, which prepares the initial
internal states as $ \bf{\hat{M}}(\pi/2,\pi)\ket{1}=1/\sqrt{2}(\ket{1}+\ket{2}) $. For the standard QWs in our experiments an additional phase offset is applied to the coin microwave pulses to cancel out a global phase that acts upon the QW due to the kicking light pulses~\cite{SiamakPRA,Siamakthesis}. A QW that has the proper phase offset is referred to as phase compensated. During the coin toss pulses a SE light
with a pulse duration of $ 30~\mu$s is added to induce well-controlled SE effect onto the QW. A delta AOKR kicking pulse
is then applied as the walk operator in momentum space followed by a coin toss microwave pulse. These coin toss pulses act
on the internal states to entangle the internal and external degrees of freedom. This sequence of coin toss followed by a
delta pulse is then repeated until QWs for $ i $ number of steps are recorded and time of flight (TOF) images are taken via the standard absorption imaging method~\cite{SiamakPRA,Siamakthesis}.

\section{Results and Discussions}
Figure~\ref{unbalanced} shows the effects of SE on a five step QW in which the phase of the walk is non-compensated
although the phase of the microwave pulses are held constant throughout the data run. The AOKR kicking strength during the
non-compensated QWs is kept at $k=1.45$, which has been proved to yield ideal QWs~\cite{SiamakPRA,Siamakthesis}. The TOF
images shown in Fig.~\ref{unbalanced}(a) indicate that the population of the atoms shifts toward the positive momentum
states as the power of the SE pulse increases. This observation confirms the prediction of Fig.~\ref{superpositionplots},
i.e., SE creates an imbalance in the internal state of the atom towards the $|F=2\rangle$ state, to a biased
momentum distribution due to the projection of that state which is moving in the direction of positive momenta. This shift
in momentum is quantitatively analyzed in Fig.~\ref{momentum}(c). In addition, the overall population of atoms present
also decreases as the SE power increases. The effective decay rate of the QW is estimated from the observed exponential
atom losses as the SE power increases (see the solid lines in Fig.~\ref{unbalanced}(b) and Fig.~\ref{unbalanced}(c)). A
typical example of our SE calibrations is shown in Fig.~\ref{unbalanced}(c) which plots the condensate fraction for a BEC
versus the duration of a single SE pulse of increasing duration at a fixed SE power of 3~$\mu$W. The BEC is first prepared
in $ F=2 $ before being subjected to a SE pulse of light increasing in $ 50~\mu s $ intervals. The exponential fitting
of this data indicates that the probability of SE event at this power is $\rho=0.35$.

We repeat the above experiment with a properly compensated QW generated at a higher kicking strength of $k\simeq 2$ to
ensure that the QW distribution is broader than those created with the lower kicking strength. We scan the SE power
up to 7.2~$\mu$W with an average of eight runs per power setting. The observed distribution of the QWs does not
show noticeable differences beyond this power value. Typical TOF images of the compensated QWs are shown in
Fig.~\ref{momentum}(d), which indicate that the atoms in the compensated walks also shift toward the positive momentum
states as the SE power increases. The decay of
the QW distribution can be more easily discerned from the momentum distributions, as displayed in Figs.~\ref{momentum}(a)
and~\ref{momentum}(b). We also extract the mean momentum and mean energy from the non-compensated and compensated QW data,
and respectively show them as a function of the SE power in Fig.~\ref{momentum}(c) and Fig.~\ref{momentum}(e). A positive
shift in the mean momentum as the SE pulses become more powerful is confirmed in Fig.~\ref{momentum}(c) for both
compensated and non-compensated QWs. In both cases the mean momentum was initially negative due to the phase of the applied microwave coin pulses being larger than $ 2 \pi $ thus causing an initial bias toward negative mean momentum. For a standard QW in our experiments this phase on the coin microwave pulses is normally below $ (2k + \pi) $ to cancel out a global phase that acts upon the QW due to the kicking light pulses~\cite{SiamakPRA,Siamakthesis}. Although this bias can adversely affect the momentum distribution of a QW that evolves in time it did not prevent the observation of the positive shift in mean momentum as the power of the SE pulses increased. This was because the same microwave coin phase was applied throughout an experiment as the SE pulse power was scanned. On the other hand, Fig.~\ref{momentum}(e) implies that the mean energy slightly
increases for the non-compensated QWs while remaining constant for the compensated QWs as the SE power increases within
the range of 0 to 7.2~$\mu$W, i.e., the SE probability is within the range of 0 to 0.85.

Our data in Fig.~\ref{unbalanced} and Fig.~\ref{momentum} indicate that the quantum to classical walk transitions happen
at around 5~$\mu$W, much lower than the maximum SE power studied in this paper. To clearly demonstrate the transition of
walks from displaying quantum to classical behaviors under the applied SE pulses, we conduct similar experiments on walks
of various numbers of steps. For each step the kicking strength is kept at $k=1.4$ to reduce the probability of extra SE
events induced by the kicking beams. The mean energy extracted from these experiments are plotted as a function of the QW
steps for various probability $\rho$ of SE events in Fig.~\ref{meanenergy}(a), which shows that the mean energy increases
with increasing number of steps at a rate $R$ which depends on the applied SE power. The rate $R$ for the zero probability
$\rho$ of SE events appears to the largest, as shown by the red markers in Fig.~\ref{meanenergy}(a). As the SE pulse is
applied and becomes more powerful, the rate $R$ gets smaller leading to an increased difference in the mean energy among
data taken under various probability $\rho$ of SE events at a high enough step, as clearly demonstrated by the 6-step and
8-step data in Fig.~\ref{meanenergy}(a). These observations qualitatively agree with the theoretical predictions derived
under similar conditions, as shown in Fig.~\ref{meanenergy}(b). Each data set in both Fig.~\ref{meanenergy}(a) and
Fig.~\ref{meanenergy}(b) are fit with a linear function and the rate $R$ are calculated and tabulated in
Fig.~\ref{meanenergy}(c). Our observations agree well with a predicted signature for quantum to classical walk
transitions, i.e., QWs have larger mean energy than classical walks at a given step because QWs distribute ballistically
while classical walks follow Gaussian distributions~\cite{PhysicalimpQW,SiamakPRA,SiamakPRL}. Our data thus indicate that
the QWs gradually transit to classical walks with less mean energy as the applied SE effect becomes powerful enough to
destroy the entanglement of the two internal spin states. Figure~\ref{meanenergy} also shows that QWs are able to maintain
their robustness when the number of steps is lower than five.

\section{Conclusions and outlook}
We have presented quantum to classical walk transitions tuned by spontaneous emissions. The SE rate is derived from the
observed atom losses during QWs. We have demonstrated that the addition of the SE light yields quantum to classical walk
transitions and leads to biased momentum distributions, which can be well explained by our numerical simulations in both
the compensated and non-compensated QWs. Our findings suggest a scheme to control the robustness of the quantum walks and
demonstrate that the effects of the SE light are intrigue. While a SE event acts as a projective measurement on the
internal spin degree of freedom, its effect on the external motion, that is the actually observed quantity, is direct by a
change of the necessary resonance conditions for the QW, but also indirect since the motion becomes biased into the
direction into which the ground state likes to move. Hence, for the center-of-mass of our atoms, SE is not a strong but
rather a weak form of quantum measurement, with the internal state acting as an ancilla that is actually strongly
measured. Many SE events will then necessary have a larger effect than just one SE event since they bias more the walk
into one direction of the external motion. Similar ideas have been put forward, e.g., in \cite{weak}. In conclusion, our
results open further possibilities of utilizing the tunable SE light to engage on the theory of measurements in
experimentally easily accessible quantum systems.

\textbf{Acknowledgments} We thank the Noble Foundation for financial support.

\end{document}